\documentclass[aps,prl,superscriptaddress,showpacs,preprint]{revtex4-1}
\usepackage[utf8]{inputenc}

\usepackage{amsmath}
\usepackage{amssymb}
\usepackage{graphics}
\usepackage{graphicx}
\usepackage{epstopdf}
\usepackage{dcolumn}% Align table columns on decimal point
\usepackage{bm}% bold math
\usepackage{xcolor}
\usepackage[marginal]{footmisc}
\usepackage[marginal]{footmisc}
\usepackage{flushend}
\usepackage{hyperref}
\usepackage{natbib}
\hypersetup{hidelinks}
\begin{document}

\title{Source coherence orchestrates nonlinear random wave revivals}
\author{Yanlin Bai}
\affiliation{Shandong Provincial Engineering and Technical Center of Light Manipulation \& Shandong Provincial Key Laboratory of Optics and Photonics Devices, School of Physics and Electronics, Shandong Normal University, Jinan 250014, China}
\affiliation{Collaborative Innovation Center of Light Manipulations and Applications, Shandong Normal University, Jinan 250358, China}
\author{Tianyu Cao}
\affiliation{Department of Electrical and Computer Engineering, Dalhousie University, Halifax, Nova Scotia, B3J 2X4, Canada}
\author{Pujuan Ma}
\affiliation{Shandong Provincial Engineering and Technical Center of Light Manipulation \& Shandong Provincial Key Laboratory of Optics and Photonics Devices, School of Physics and Electronics, Shandong Normal University, Jinan 250014, China}
\affiliation{Collaborative Innovation Center of Light Manipulations and Applications, Shandong Normal University, Jinan 250358, China}
\author{Xin Liu}
\email[]{xinliu@sdnu.edu.cn}
\affiliation{Shandong Provincial Engineering and Technical Center of Light Manipulation \& Shandong Provincial Key Laboratory of Optics and Photonics Devices, School of Physics and Electronics, Shandong Normal University, Jinan 250014, China}
\affiliation{Collaborative Innovation Center of Light Manipulations and Applications, Shandong Normal University, Jinan 250358, China}
\author{Yangjian Cai}
\email[]{yangjian\_cai@163.com}
\affiliation{Shandong Provincial Engineering and Technical Center of Light Manipulation \& Shandong Provincial Key Laboratory of Optics and Photonics Devices, School of Physics and Electronics, Shandong Normal University, Jinan 250014, China}
\affiliation{Collaborative Innovation Center of Light Manipulations and Applications, Shandong Normal University, Jinan 250358, China}
\author{Sergey A. Ponomarenko}
\email[]{serpo@dal.ca}
\affiliation{Department of Electrical and Computer Engineering, Dalhousie University, Halifax, Nova Scotia, B3J 2X4, Canada}
\affiliation{Department of Physics and Atmospheric Science, Dalhousie University, Halifax, Nova Scotia, B3H 4R2, Canada}
\author{Chunhao Liang}
\email[]{chunhaoliang@sdnu.edu.cn}
\affiliation{Shandong Provincial Engineering and Technical Center of Light Manipulation \& Shandong Provincial Key Laboratory of Optics and Photonics Devices, School of Physics and Electronics, Shandong Normal University, Jinan 250014, China}
\affiliation{Collaborative Innovation Center of Light Manipulations and Applications, Shandong Normal University, Jinan 250358, China}

\date{\today}

\begin{abstract}
We demonstrate that the Talbot length of periodic wave packets, long believed to be solely determined by their periodicity, is strongly affected by the source coherence in the nonlinear propagation regime. We reveal that reducing the source coherence---and consequently the speckle size of a periodic field---shortens the Talbot length and significantly improves the quality of the recurrent Talbot images, despite a fixed source periodicity. This effect arises from coherence-mediated nonlinear mode coupling, which alters a phase synchronization condition undergirding the wave packet revivals. Our findings expose a hidden role of coherence in governing Talbot revivals of random waves in the nonlinear regime, crucially informing the understanding and application of the Talbot effect in realistic media.

\end{abstract}
\maketitle
\textit{Introduction---} The Talbot effect is a fundamental self-imaging phenomenon in wave physics that manifests the self-reconstruction of a periodic wave field during paraxial propagation~\cite{wen2013talbot}. Such Talbot revivals occur in specific transverse planes known as Talbot planes. Originally observed in diffraction of light from a grating~\cite{talbot1836facts}, the Talbot effect arises from  constructive interference of discrete spatial (temporal) frequency components that accumulate deterministic phase shifts on propagation~\cite{rayleigh1881xxv}. As a result, the initial periodic structure reappears at integer multiples of the Talbot distance, while fractional self-images emerge at rational fractions of this distance~\cite{Berry1996integer}. Recently, more general types of Talbot revivals have been discovered for spatiotemporally structured light wave packets~\cite{yessenov2020veiled,hall2021space}. In addition, the mathematical analogy between the paraxial wave equation for classical light and the Schr\"{o}dinger equation ensures the existence of quantum wave revivals that have been widely explored~\cite{parker1986coherence, averbukh1989fractional,loinaz1999quantum,banerji2007exploring,song2011experimental}, especially in the context of quantum carpets~\cite{berry2001quantum,saif2001quantum,kazemi2013quantum}, and found numerous applications~\cite{farias2015quantum,barros2017free}. Due to their universal nature, wave packet revivals have been observed in a wide variety of physical systems, including optical fields~\cite{patorski1989self}, matter waves~\cite{deng1999temporal,ryu2006High}, and even water waves~\cite{rozenman2022periodic,bakman2019observation}. Beyond their fundamental importance, such revivals have also enabled numerous applications in imaging~\cite{wen2013talbot,han2013wide,Pfeiffer2008Xray}, prime number decomposition~\cite{bigourd2008factorization,pelka2018prime,liu2023axial}, and waveguide arrays~\cite{Iwanow2005Discrete}.

Much of the work on the Talbot effect and wave packet revivals in general has been concerned with fully coherent wave packets evolving in free space or linear media. In this context, the self-imaging mechanism can be understood as a linear superposition of quadratic phases of the discrete angular spectrum of plane waves~\cite{wen2013talbot}. Under these conditions, both integer and fractional revivals arise naturally, and the recurrence distance is determined solely by the diffraction or dispersion properties of the medium. However, as the medium nonlinearity is introduced, the underlying physics changes drastically~\cite{zhang2014Nonlinear, schiek2021excitation,rozenman2022periodic}. In particular, the wave evolution in nonlinear dispersive systems, described within the framework of the nonlinear Schrödinger equation (NLSE), is governed by the interplay between dispersion and Kerr nonlinearity. In this regime, the revivals are no longer a simple consequence of linear interference, but are instead the result of nonlinear interactions of localized wave structures. Previous studies have shown that periodic coherent nonlinear waves, such as Akhmediev breathers, can exhibit nonlinear revivals with the Talbot carpet significantly differing from the linear one: fractional revivals completely disappear, leaving only recurrences over integer and half-integer multiples of the revival distance~\cite{zhang2014Nonlinear}. 

However, there is unavoidable noise in any realistic nonlinear wave system, caused by either light source fluctuations or medium parameter fluctuations. This raises a series of fundamental questions: Can wave packet revivals occur in nonlinear media in the presence of even a modest amount of noise? If so, how do these fluctuations, which degrade the spatial and/or temporal coherence of the propagating wavefields,  affect the quality of primary and recurrent images, the magnitude of the revival distance, and the Talbot carpet structure? Addressing these fundamental questions will critically advance our understanding of the physics of nonlinear random wave revivals and crucially inform Talbot effect applications in areas such as lensless imaging, information encoding and transfer, optical metrology, and spectroscopy in realistic nonlinear media.

Importantly, wave packet revivals have been studied in the context of random or partially coherent fields, including self-imaging of random wave packets in linear graded-index media~\cite{ponomarenko2015self}, Talbot revivals of partially coherent beams upon reflection by a grating~\cite{schouten2025gratings}, and revivals of quasi-periodic random waves in linear media~\cite{Bai2026Regular}. However, all of these studies have been limited to wave packet recurrences during their evolution in \emph{strictly linear systems}. Therefore, the exploration of Talbot revivals of random waves in nonlinear systems remains an uncharted frontier.

Here, we address this open problem by developing a theoretical model for periodic partially coherent wave packet evolution in Kerr nonlinear media. To this end, we consider an ensemble of random wave packets with periodic deterministic envelopes and multiplicative noise with periodic second-order correlations. A physical realization of such a wave packet can be furnished by an optical coherence lattice (OCL)~\cite{ma2014optical}. To date, such OCLs have been extensively studied theoretically~\cite{OCL15T,OCL17T} and realized in the laboratory~\cite{OCL16E,OCL21E}. We discover that the coherence of the source of such a structured OCL governs Talbot-like recurrences. Specifically, we show that coherence acts as a crucial control parameter of nonlinear Talbot dynamics, enabling the control of not only the revival distance, but also the fine structure of the Talbot carpet and the wave packet self-reconstruction quality, without altering either the physical properties of the medium, the source power, or the period of the speckled source. The ability to tune Talbot recurrences through coherence engineering not only provides new insight into the fundamental physics of nonlinear wave propagation but also suggests potential strategies for controlling recurrences of wave packets in nonlinear media.

\textit{Theory---} We consider a train of random pulses,  propagating in a self-focusing Kerr nonlinear medium. The pulse evolution in the medium is governed by the celebrated nonlinear Schrödinger equation (NLSE) of the form
\begin{equation}\label{1}
i \frac{\partial U}{\partial z} + \frac{\beta_2}{2} \frac{\partial^2 U}{\partial t'^2} + \gamma |U|^2 U = 0,
\end{equation}
where \( t' = t - z/v_g \) is the retarded time in the reference frame co-propagating with the pulse group velocity \( v_g \), \( z \) is the propagation distance along the medium, \( U(z,t') \) denotes the slowly varying envelope of the optical pulse, \( \beta_2 \) (\(<0\)) is the second-order group velocity dispersion (GVD) parameter, and \( \gamma \) is the Kerr nonlinear coefficient.

Let us now describe a speckled light source. To this end, we take the field at the source to be a product of a complex random amplitude $\alpha (t)$ and a deterministic envelope $\psi (t)$, such that 
\begin{equation}\label{2}
    U_0 (t)=U(t, z = 0) = \alpha(t)\psi(t),
\end{equation}
and assume $\psi(t)$ and $\alpha(t)$ to be periodic with the same period $T$. We note in passing that although we have recently discovered profound implications of commensurability or noncommensurability between the intensity and its degree-of-coherence periods at the source for linear Talbot carpets~\cite{Bai2026Regular}, here we focus on the interplay of source coherence and medium nonlinearity, and hence we keep the two periods the same. 

Next, exploiting its periodicity, $\alpha(t)$ can be decomposed into a Fourier series as
\begin{equation}\label{3}
\alpha(t) = \sum_m a_m e^{-i 2\pi m t / T},
\end{equation}
where in all summations the indices run from $-\infty$ to $\infty$ hereafter; \( \{a_m\} \) are uncorrelated, statistically stationary random Fourier amplitudes, obeying the following statistics:
\begin{equation}\label{4}
\left\langle a_m^* a_n \right\rangle = \lambda_m \delta_{mn}.
\end{equation}
Here the asterisk denotes complex conjugation and the angular brackets stand for ensemble averaging; \( \delta_{mn} \) is the Kronecker delta function. We can express random Fourier amplitudes as  $a_m = \sqrt{\lambda_m} e^{i \phi_m}$, where $\{\phi_m\}$ is a set of random phases uniformly distributed over the interval $(-\pi, \pi]$.

Using Eqs.~(\ref{2}) through~(\ref{4}), we can express the second-order mutual coherence function of the source field, defined as \( \Gamma(t_1, t_2) = \left< U_0^*(t_1) U_0(t_2) \right> \), in the form 
\begin{equation}\label{5}
\Gamma(t_1, t_2) = \psi_0^*(t_1) \psi_0(t_2) \mu(t_1 - t_2).
\end{equation}
Equation \eqref{5} describes an OCL~\cite{ma2014optical} with a periodic degree of coherence as
\begin{equation}\label{6}
\mu(t_1,t_2) = \sum_m \lambda_m e^{-i 2\pi m (t_2 - t_1)/T}.
\end{equation}
Each OCL of Eq.~(\ref{6}) is determined by a distribution of the set $\{\lambda_m\}$ of modal weights. We assume a Gaussian distribution given by
\begin{equation}\label{7}
\lambda_m = \lambda_0 e^{-2 \xi_c m^2},
\end{equation}
where $\xi_c$ is a normalized coherence parameter that describes the global state of coherence of the OCL, and $\lambda_0$ is a normalization constant. In the limit \( \xi_c \to \infty \), the source is fully coherent, being composed of the lowest-order \( m=0 \) mode. In this limit, the degree of coherence approaches unity, which is the hallmark of a fully coherent state~\cite{PSA05}. In the limit \( \xi_c \to 0 \), we have a nearly incoherent source composed of a multitude of the Fourier modes of nearly equal weights. 

Furthermore, we assume the deterministic pulse envelope \( \psi_0(t) \) to be an Akhmediev breather (AB)~\cite{Akhmediev1987exact,dudley2009modulation}, which is an exact periodic solution of the NLSE. At $z=0$, it is given by
\begin{equation}\label{AB}
\psi_0(t) = \sqrt{P_0}\left[\frac{1 - 4a + \sqrt{2a} \cos(\omega t)}{\sqrt{2a} \cos(\omega t) - 1}\right],
\end{equation}
where \( \omega = 2\pi / T \) is the angular frequency of the envelope. We can express the AB period in terms the power $P_0$ of any of its peaks and the modulation parameter \( a \) as
\begin{gather}
T=\frac{\pi\sqrt{|\beta_2|}}{\sqrt{\gamma P_0}\sqrt{1-2a}}, \qquad 0 < a < 1/2.
\end{gather}
Here $a$ determines the AB periodicity for a given $P_0$; in the limit $a\rightarrow 1/2$, the AB turns into an aperiodic Peregrine soliton~\cite{dudley2009modulation}. We can expand the AB amplitude of Eq.~(\ref{AB}) into a Fourier series, with coefficients $\{ c_n\}$ calculated in the Supplemental Material~\cite{SI}, yielding
\begin{equation}\label{9}
    U_0(t)=\sum_{n,m}a_m c_ne^{-i2\pi (n+m)t/T}.
\end{equation}

To numerically simulate the evolution of the pulse train ensemble, it is convenient to transform to dimensionless variables, $\tau=t/\sigma_I$, $\zeta=z/L_D$, and $\Psi=U/\sqrt{P_0}$. Here $\sigma_I$ is a characteristic width of individual peak of the AB breather; it is determined by the peak power $P_0$ and the modulation parameter $a$. We can then introduce two fundamental characteristic lengths of the NLSE, the nonlinear length \( L_{\text{NL}} \) and the dispersion length \( L_D \), as
\begin{equation}\label{L}
L_{\text{NL}} = \frac{1}{\gamma P_0}, \quad L_D = \frac{\sigma_I^2}{|\beta_2|}.
\end{equation}
In the dimensionless variables, the NLSE takes the form
\begin{equation}\label{11}
i\frac{{\partial \Psi }}{{\partial \zeta }} + \frac{1}{2}\frac{{{\partial ^2}\Psi }}{{\partial {\tau ^2}}} + {N^2}{\left| \Psi  \right|^2}\Psi  = 0,
\end{equation}
where $N = \sqrt{ {L_D}/{L_{NL}}}$ is a soliton parameter that governs the pulse dynamics in a Kerr medium; the larger the $N$, the stronger the effect of the medium nonlinearity.

\begin{figure}[t]
\centering
\includegraphics[width = 0.85\linewidth]{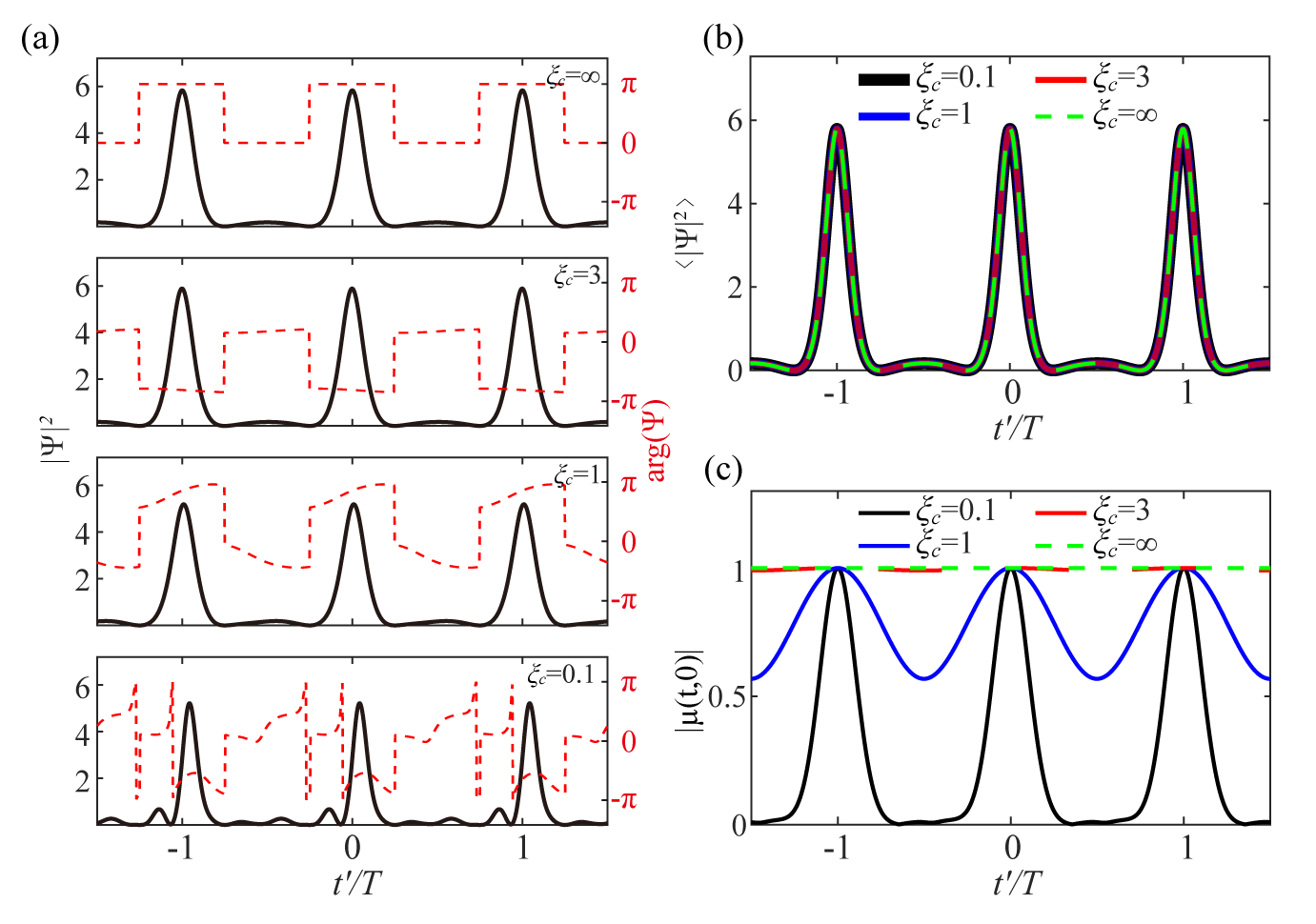}
\caption{Ensemble realizations of a speckled pulse train at the source and the corresponding ensemble averaged intensity and degree of coherence. (a) Intensity and phase distributions of an ensemble realization for variable source coherence parameter $\xi_c$. The lower the $\xi_c$, the greater the randomness and the smaller the typical speckle size at the source. (b) Ensemble-averaged intensity distribution at the source. (c) Ensemble-averaged degree of coherence at the OCL-like source. The degree of coherence at the source flattens out as $\xi_c$ increases.} 
\label{fig1}
\end{figure}

\textit{Results and discussion---}Let us describe an ensemble of our speckled pulse sources. In Fig.~\ref{fig1}(a), we display the intensity (solid black curves) and phase (red dash curves) of random realizations of the source ensemble for a fixed modulation parameter $a = 0.25$ and variable global coherence $\xi_c$. In the limit $\xi_c=\infty$, only the fundamental mode survives, and the random amplitude is given by $\alpha(t)=a_0$ (see Eq.~(\ref{3})). Hence, the field is fully determined by the deterministic AB amplitude, yielding a smooth intensity profile with phases restricted to 0 or $\pi$ (top row of Fig.~\ref{fig1}(a)). Phase singularities coincide with intensity zeros. As the global coherence $\xi_c$ decreases, higher-order modes enter the Fourier decomposition of $\alpha (t)$, resulting in increasingly pronounced fluctuations in both the intensity and phase of the field at the source (second and third rows from the top). In particular, the source field corresponding to $\xi_c=0.1$ is quite noisy. Notably, the global coherence reduction does not alter the periodicity of the wave packets generated by the source. Rather, the intensity and phase of each pulse within the train are modified with $\xi_c$. We stress that the figure serves to illustrate the qualitative behavior of representative individual realizations for different states of source coherence. Quantitatively, the global degree of coherence of the source is set by the value of $\xi_c$.

Each random wave packet represents an individual realization of the source ensemble. By refreshing the random phases $\{\phi_m\}$ of Fourier modes, we can generate new realizations. Statistical stationarity can be attained when a sufficiently large ensemble is collected. The resulting partially coherent wave packet, shown in Fig.~\ref{fig1}(b) and ~\ref{fig1}(c), retains the average intensity while the degree of coherence of the ensemble gradually evolves from unity for fully coherent light to a structured OCL with a periodic pattern (see Eq.~(\ref{6})). All individual OCL peaks narrow as the source coherence decreases. Importantly, for any partially coherent wave packet, both the intensity and the degree of coherence exhibit periodicity with a fixed period, independent of $\xi_c$. As each realization of the pulse train ensemble propagates in a linear dispersive medium, the Talbot distance, determined by $z_T=T^2/\pi|\beta_2|$~\cite{patorski1989self}, depends solely on the period of the train. In our case, the dimensionless linear Talbot distance equals $Z_T\approx6.28$.

\begin{figure}[t]
\centering
\includegraphics[width =0.7\linewidth]{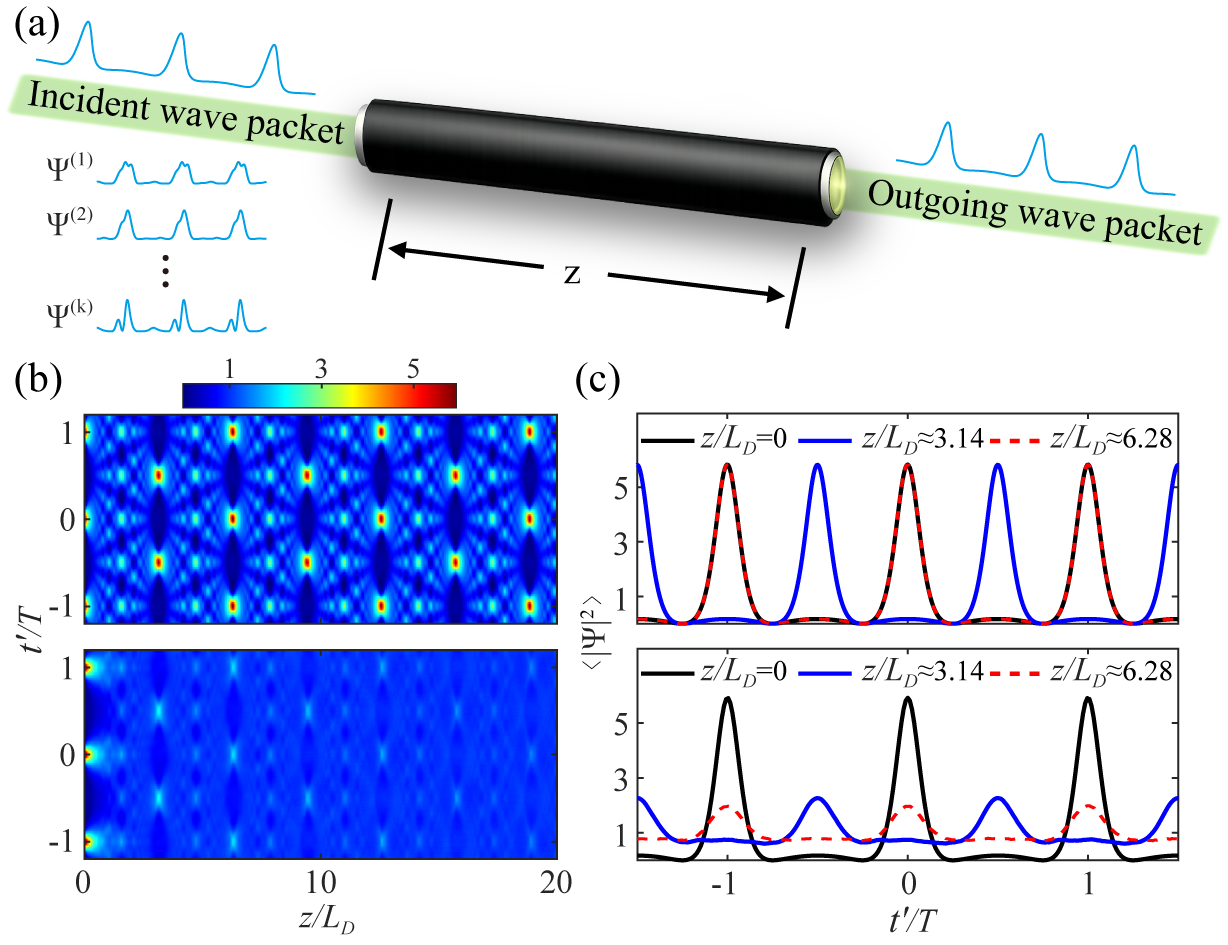}
\caption{(a) Sketch of revivals of random wave packets in a nonlinear dispersive medium, such as an optical fiber. (b) Talbot carpets generated by a low-coherence source ($\xi_c=0.1$) in  a linear dispersive medium (top panel) and a nonlinear dispersive medium (bottom panel). (c) Ensemble-averaged intensity profiles at selected propagation distances corresponding to the panels in (b).} 
\label{fig2}
\end{figure}

To study the Talbot dynamics numerically in the nonlinear propagation regime, we consider an ensemble of random wave packets propagating through a nonlinear dispersive medium, such as an optical fiber (see Fig.~\ref{fig2}(a)). We can obtain the intensity $I(\tau,\zeta)=|\Psi(\tau,\zeta)|^2$ at distance $\zeta$ from the source by averaging a large number of realizations, 
\begin{equation}\label{ave}
    I(\tau,\zeta)\approx\frac{1}{M}\sum_{k=1}^{M}|\Psi^{(k)}(\tau,\zeta)|^2.
\end{equation}
If the number of realizations $M$ is large enough, Eq.~(\ref{ave}) yields a good approximation for the ensemble average. By monitoring the evolution of the intensity along the propagation distance, we analyze the emergence of wave packet revivals. For each value of $\xi_c$, we identify the recurrence distance from the prominent local maxima of the axial intensity signal $I(\tau=0,\zeta)$. We propagate each realization numerically by solving the NLSE with a standard split-step Fourier method~\cite{Agrawal2012Nonlinear}. 

To understand the role of source coherence and nonlinearity in wave packet revivals, we note that a phase synchronization condition is crucial for revivals to occur. In a nonlinear medium, this condition is dictated by the interaction among periodic nonlinear eigenmodes of the system mediated by noise. In the low-coherence regime, $\xi_c\ll 1$, dispersion dominates the dynamics of the system, leading to eigenmodes that remain close to plane waves. It follows that the Talbot length is largely independent of the source coherence and is determined by the source period. However, at the same time, the nonlinearity impedes phase synchronization of the eigenmodes, thereby degrading the quality of wave packet revivals. Specifically, the nonlinearity causes blurring of the fine features of the Talbot carpet and reduction of the peak amplitudes of the source field intensity. In Fig.~\ref{fig2}(b), we illustrate this effect by comparing Talbot carpets produced by the same relatively incoherent source with $\xi_c=0.1$ in  linear (top panel) and nonlinear (bottom panel) media, respectively. The top panel of Fig.~\ref{fig2}(b) shows a Talbot carpet with clearly visible fine structure, while the latter is discernible but significantly degraded in the bottom panel of the figure. We notice that the intensity peaks of the periodic source images are substantially attenuated (see Fig.~\ref{fig2}(c)).

\begin{figure}[t]
\centering
\includegraphics[width =0.7\linewidth]{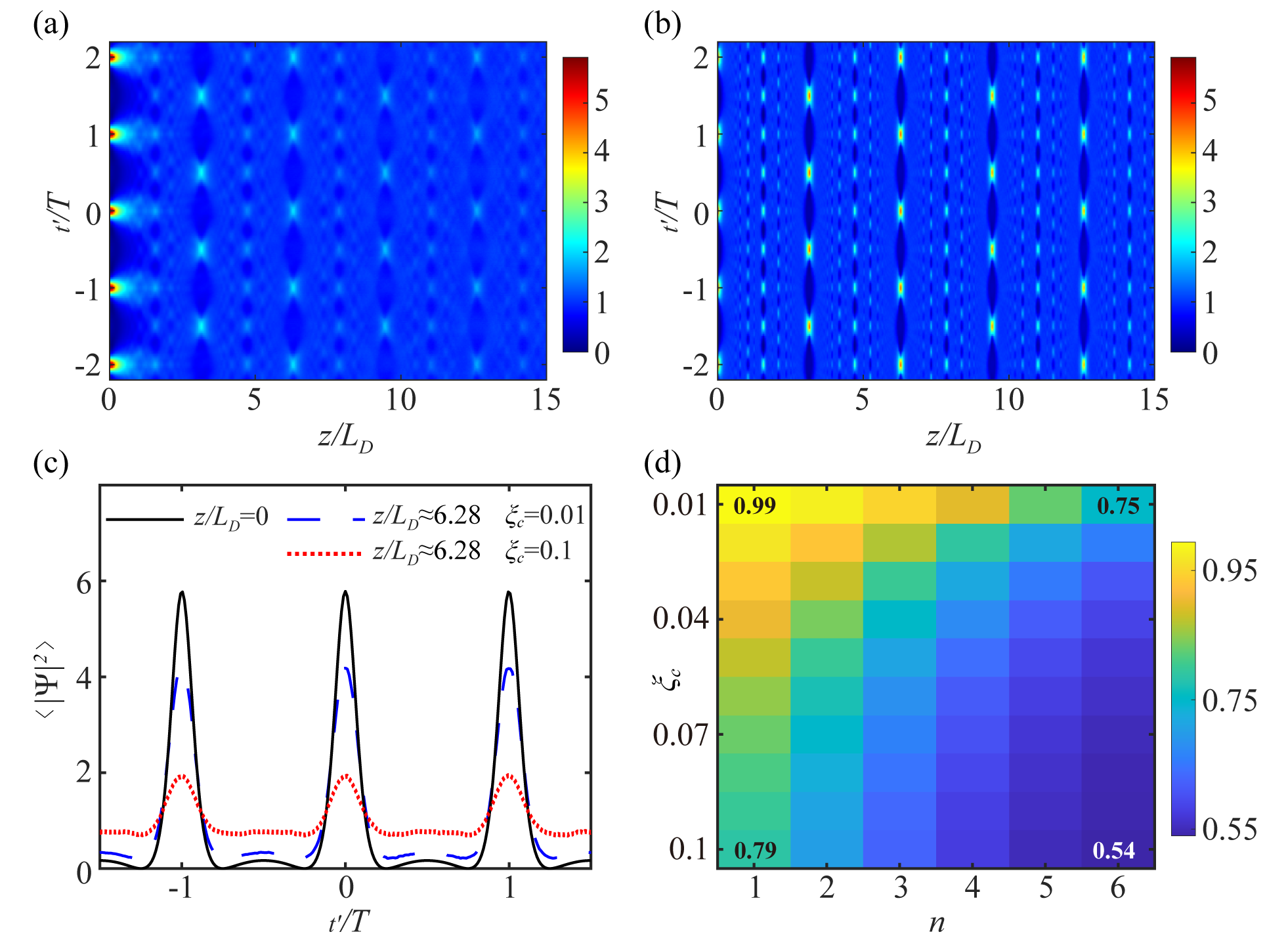}
\caption{Talbot carpet of an ensemble of wave packets of low coherence in a nonlinear dispersive medium: (a) $\xi_c=0.1$, (b) $\xi_c=0.01$. (c) Intensity profiles at the first Talbot revival distance for different coherence parameters, compared with the source field. (d) Image fidelity $F$ as a function of the recurrence order $n$ and the source coherence parameter $\xi_c$.}
\label{fig3}
\end{figure}

As the source coherence decreases further towards the nearly incoherent limit, $\xi_c\rightarrow 0$, a complex Talbot carpet, replete with conspicuous fine features reemerges, as is evident from comparing Figs.~\ref{fig3}(a) and (b). In Fig.~\ref{fig3}(c), we show the intensity profiles at the first recurrence distance for the two source coherence parameters, together with the source intensity profile. For $\xi_c=0.01$, the reconstructed intensity profile almost overlaps with the source profile, indicating a high-quality Talbot-like revival. In contrast, for $\xi_c=0.1$, the reconstructed profile exhibits noticeable deviations from that at the source. It follows that the source coherence reduction suppresses nonlinear distortions and improves the overall self-reconstruction quality of the wave packet. In addition, higher-order recurrences can be identified for the wave packets generated by low-coherence sources as is evidenced in Fig.~\ref{fig3}(d).

To explain these results, we notice that reducing the source coherence does not alter either the source period or the GVD parameter $\beta_2$, leaving the conventional (linear) Talbot length unaffected. Instead, the reduction in source coherence introduces smaller effective speckle scales, which fundamentally alters the balance between local dispersion and Kerr nonlinear mode coupling. Indeed, in the low-coherence limit, the effective pulse width $\sigma_I'$ is essentially independent of the width of an AB peak. Rather, the pulse width is determined by the effective size of a random wave field speckle. The speckle size shrinks as the source coherence diminishes [see Fig.~\ref{fig1}(a)]. We can estimate the speckle size as $\sigma_I'\sim \xi_c \sigma_I$. The corresponding effective dispersion length $L_D'=\sigma_I'^2/|\beta_2|$ is then reduced by roughly a factor of $\xi_c^2$ (see Eq.~(\ref{L})). It follows that the wave packet evolution is essentially linear, with $L_D'\ll L_{NL}$~\cite{bai2025Self,chen2025Controllable}, and the eigenmodes are virtually indistinguishable from plane waves. Such eigenmodes can satisfy the phase-synchronization condition nearly perfectly, thereby generating a complex Talbot carpet, almost identical to the one for the linear case~\cite{Bai2026Regular}.  In this context, we can quantify the quality of self-imaging by a fidelity figure-of-merit $F$, which we define as 
\begin{equation}
F(\zeta,\xi_c) = \sup_{\tau \in \mathbb{R}} \frac{\int d\tau' I_0(\tau',\xi_c) I (\tau + \tau', \zeta,\xi_c)}{\sqrt{\int d\tau I_0^2(\tau,\xi_c)} \sqrt{\int d\tau I^2(\tau, \zeta,\xi_c)}},
\end{equation}
where $I_0$ is the intensity of the source and ``sup" denotes the supremum over all possible $\tau\in\mathbb{R}$ for given values of $\zeta$ and $\xi_c$. 

To examine the revival stability, we evaluate the fidelity for successive recurrence orders. Figure~\ref{fig3}(d) shows the revival fidelity $F$ as a function of the recurrence order $n$ for different source coherence parameters $\xi_c$. For $\xi_c=0.01$, the primary image fidelity reaches $F=0.99$. Furthermore, the fidelity of even the sixth-order recurrence remains as high as $F=0.75$, indicating that the low-coherence recurrences emerge at the multiples of the revival length. By contrast, for $\xi_c=0.1$, the fidelity decreases from $F=0.79$ for the primary image to $F=0.54$ for the sixth-order one. This rapid image quality degradation can be attributed to the accumulation of residual nonlinear phase mismatch, which progressively perturbs the phase-synchronization condition required for wave packet revivals. Thus, reducing the source coherence not only improves the initial reconstruction quality, but also enhances the long-distance stability of recurrences.

\begin{figure}[t]
\centering
\includegraphics[width =0.7\linewidth]{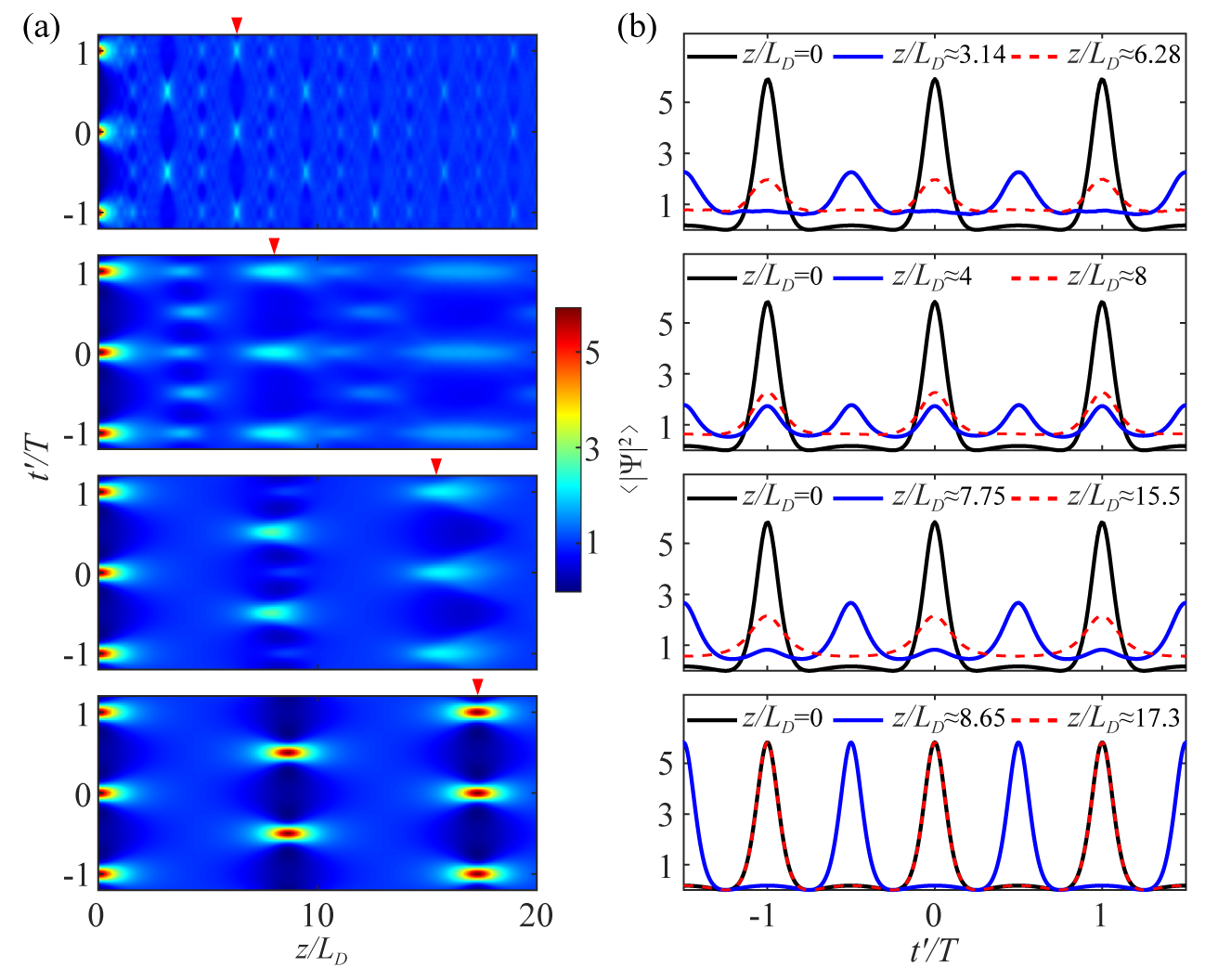}
\caption{(a) Talbot carpet of an ensemble of random wave packets in a nonlinear dispersive medium for variable source coherence (top to bottom): \(\xi_c=0.1\), \(\xi_c=4\), \(\xi_c=8\), and fully coherent. (b) Ensemble-averaged intensity profiles at selected propagation distances for the corresponding coherence parameters.} 
\label{fig4}
\end{figure}

However, as the source coherence increases outside of the low-coherence regime, the Talbot carpet becomes progressively simpler, shedding off its fine structure, until bright, sharp self-images of the source emerge in the high-coherence regime. We illustrate this evolution in Fig.~\ref{fig4}(a), with the corresponding intensity profiles shown in Fig.~\ref{fig4}(b). To explain these dynamics we note that as the global coherence of the source increases, the random ABs become progressively less noisy. For sufficiently large $\xi_c$, the initial noise contributes minimally to the source field, making it nearly identical to a deterministic AB. The latter is a periodic nonlinear eigenmode of the NLSE, known to yield a simple Talbot carpet~\cite{zhang2014Nonlinear}. 

\begin{figure}[t]
\centering
\includegraphics[width =0.5\linewidth]{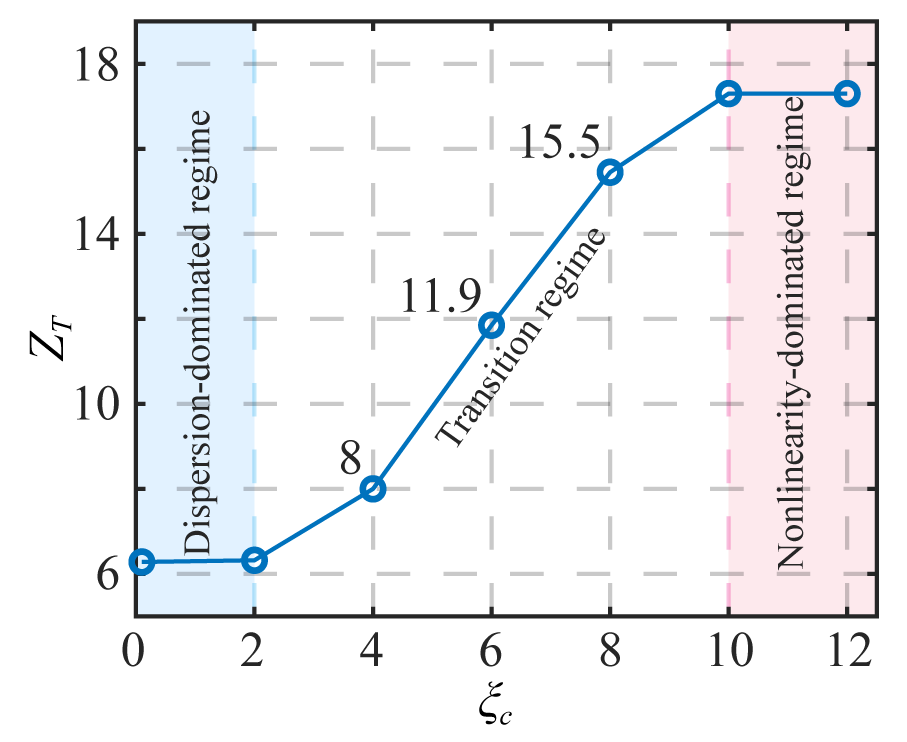}
\caption{Dimensionless Talbot distance $Z_T$ as a function of the coherence parameter \( \xi_c \), with background colors distinguishing dispersion-dominated (light blue), transition, and nonlinearity-dominated (pink) regime.}
\label{fig5}
\end{figure}

At the same time, it is instructive to monitor the dependence of the Talbot length on the source coherence. In Fig.~\ref{fig5}, we exhibit the Talbot distance as a function of the coherence parameter $\xi_c$. We can infer from the figure that following an initial plateau in the low-coherence regime, the Talbot distance monotonically grows with $\xi_c$, eventually saturating at large enough values of the source coherence parameter. In the low-coherence region, highlighted in light blue, the dispersion effects dominate. The resulting eigenmodes are very close to plane waves for which the phase-synchronization condition does not depend on source coherence~\cite{Bai2026Regular}. By the same token, in the high-coherence region, highlighted in light pink, the nonlinear eigenmodes are nearly periodic and their period, rather than negligible source noise, determines the Talbot distance. The monotonic growth of the Talbot length in the intermediate region of $\xi_c$ can be explained as follows. If the initial condition is not a nonlinear eigenmode of the system, the eigenmodes form only over a finite propagation distance away from the source. This is the case for an AB breather with OCL-like correlations. Therefore, nonlinear random wave packets self-reconstruct in general over longer Talbot distances than do their linear cousins. At the same time, lower noise at the source (larger $\xi_c$) leads to a greater role of nonlinear mode coupling in phase synchronization, which implies that wave packet revivals occur over longer distances. We determine the revival distance in the intermediate (partially coherent) region by examining the peaks of the ensemble-averaged axial intensity and quantitatively assess the image quality by evaluating $F$. In this cross-over region, the eigenmodes of the system are neither plane waves nor pure Akhmediev breathers. Therefore, the phase synchronization condition cannot be analytically derived and the wave recurrences are governed by an intricate interplay of nonlinearity and dispersion mediated by noise. Further research is required to provide a comprehensive quantitative description of the recurrences in this cross-over region.

Finally, we would like to distinguish the subtle nuances in how the interplay of source coherence and medium nonlinearity affect the Talbot carpet structure and the Talbot distance. In the low-coherence limit, the Talbot distance is essentially independent of source coherence, whereas fine features of the Talbot carpet undergo blurring due to weak nonlinear mode coupling, followed by total disappearance. As the source coherence further increases beyond the low-coherence limit, a simple Talbot carpet forms with progressively sharper features. To illustrate this point, we show Talbot carpet formation and subsequent evolution as the source coherence increases in Supplemental Video 1~\cite{SI}. We observe that for $\xi_c \in [0.3, 1.3]$ the Talbot carpet features are quite blurred, while the Talbot distance remains essentially the same (see Fig.~\ref{fig5}). By the same token, we observe how the Talbot carpet transforms, losing its fine structure for $\xi_c>1.3$. For $\xi_c\simeq 10$, it eventually approaches the pattern typical of a fully coherent Talbot carpet in a nonlinear medium~\cite{zhang2014Nonlinear}.

%\begin{figure}[t]
%\centering
%\includegraphics[width =0.5\linewidth]{6.png}
%\caption{Dimensionless Talbot distance $Z_T$ as a function of the dimensionless source period \(T/T_0\) for a variable coherence parameter \(\xi_c\).} 
%\label{fig6}
%\end{figure}

%This raises the question of whether such a trend persists across different source periods. Fig.~\ref{fig6} addresses this by plotting the revival distance as a function of \(T\) for several coherence parameters. It is evident that the overall decreasing trend with coherence holds for all \(T\), yet the sensitivity to \(T\) itself is gradually suppressed as coherence diminishes, ultimately converging to the linear Talbot limit.

\textit{Conclusions---}
We have shown that the global coherence of a periodic source plays a fundamental role in governing Talbot revivals of random wave packets generated by the source and propagating in a Kerr-like nonlinear medium. Specifically, we have demonstrated that not only the Talbot revival length, but also the fine structure of the Talbot carpet are fundamentally affected by the source coherence. Our results imply that the global coherence of the source can serve as a convenient control knob to suppress coherence-mediated nonlinear mode coupling---and thereby alter the phase synchronization condition underlying wave packet revivals---without affecting the power of the source. As a result, high-intensity wave packets in nonlinear media can feature fractional revivals and the Talbot length dependence on the period of the source field, characteristic of the Talbot effect in free space or linear dispersive media. Thus, our results reveal a new mechanism of coherence-controlled nonlinear wave dynamics and provide a route to tailoring self-imaging phenomena in complex nonlinear systems. Besides its inherent fundamental importance, our work can find applications in optical imaging and information transmission with random light through nonlinear media.

%\textcolor{blue}{Finally, some speculations on individual ensemble member revivals. As individual ensemble representations can vary in peak amplitudes, their revival distances can also vary in the nonlinear regime. This is because the propagation distance over which the bona fide system eigenmodes form depends on the peak power of the source. However, averaging over individual realizations translates to averaging over the revival distances, yielding a somewhat lower quality (blurred) replica of the mutual coherence function of the source at a single Talbot distance determined, in general, by the soliton parameter, source coherence, and the period of the source.}

\textit{Acknowledgments---}The authors acknowledge financial support from National Key Research and Development Program of China (2022YFA1404800); National Natural Science Foundation of China (12534014, 12374311, 12192254, W2441005); Taishan Scholars Program of Shandong Province (tsqn202312163); Natural Science Foundation of Shandong Province (ZR2025ZD21,ZR2023YQ006); Key research and development program of Shandong Province (2024JMRH0105) and the Natural Sciences and Engineering Research Council of Canada (RGPIN-2025-04064). 

\textit{Data availability---}The data that support the findings of
this article are not publicly available. The data are available
from the authors upon reasonable request.

%\bibliography{references}

\begin{thebibliography}{99}

\bibitem{wen2013talbot}
J.~Wen, Y.~Zhang, and M.~Xiao,
The Talbot effect: recent advances in classical optics, nonlinear optics, and quantum optics,
Adv. Opt. Photon. \textbf{5}, 83 (2013).

\bibitem{talbot1836facts}
H.~F.~Talbot,
Facts relating to optical science,
Philos. Mag. \textbf{9}, 401 (1836).

\bibitem{rayleigh1881xxv}
Lord Rayleigh,
XXV. On copying diffraction-gratings, and on some phenomena connected therewith,
Philos. Mag. \textbf{11}, 196 (1881).

\bibitem{Berry1996integer}
M.~V.~Berry and S.~Klein,
Integer, fractional and fractal Talbot effects,
J. Mod. Opt. \textbf{43}, 2139 (1996).

\bibitem{yessenov2020veiled}
M.~Yessenov, L.~A.~Hall, S.~A.~Ponomarenko, and A.~F.~Abouraddy,
Veiled Talbot effect,
Phys. Rev. Lett. \textbf{125}, 243901 (2020).

\bibitem{hall2021space}
L.~A.~Hall, M.~Yessenov, S.~A.~Ponomarenko, and A.~F.~Abouraddy,
The space--time Talbot effect,
APL Photon. \textbf{6}, 050801 (2021).

\bibitem{parker1986coherence}
J. Parker and C.~R.~Stroud Jr,
Coherence and decay of Rydberg wave packets,
Phys. Rev. Lett. \textbf{56}, 716 (1986).

\bibitem{averbukh1989fractional}
I.~Sh.~Averbukh and N.~F.~Perelman,
Fractional regenerations of wave packets in the course of long-term evolution of highly excited quantum systems,
Sov. Phys. JETP \textbf{69}, 464 (1989).

\bibitem{loinaz1999quantum}
W. Loinaz and T.~J.~Newman,
Quantum revivals and carpets in some exactly solvable systems,
J. Phys. A: Math. Gen. \textbf{32}, 8889 (1999).

\bibitem{banerji2007exploring}
J.~Banerji,
Exploring fractional revivals and sub-Planck structures: a walk in phase space with Wigner and Kirkwood,
Contemp. Phys. \textbf{48}, 157 (2007).

\bibitem{song2011experimental}
X-B. Song, H-B. Wang, J. Xiong, K. Wang, X. Zhang, K-H. Luo, and L-A. Wu,
Experimental observation of quantum Talbot effects,
Phys. Rev. Lett. \textbf{107}, 033902 (2011).

\bibitem{berry2001quantum}
M. Berry, I. Marzoli, and W. Schleich,
Quantum carpets, carpets of light,
Phys. World \textbf{14}, 39 (2001).

\bibitem{saif2001quantum}
F. Saif and M. Fortiano,
Quantum revivals in periodically driven systems close to nonlinear resonances,
Phys. Rev. A \textbf{65}, 013401 (2001).

\bibitem{kazemi2013quantum}
P.~Kazemi, S.~Chaturvedi, Irene Marzoli, R.~F.~O'Connell, and W.~P.~Schleich,
Quantum carpets: a tool to observe decoherence,
New J. Phys. \textbf{15}, 013052 (2013).

\bibitem{farias2015quantum}
O. J. Farías, F. De Melo, P. Milman, and S. P. Walborn,
Quantum information processing by weaving quantum Talbot carpets,
Phys. Rev. A \textbf{91}, 062328 (2015).

\bibitem{barros2017free}
M. R. Barros, A. Ketterer, O. J. Farías, and S. P. Walborn,
Free-space entangled quantum carpets,
Phys. Rev. A \textbf{95}, 042311 (2017).

\bibitem{patorski1989self}
K.~Patorski,
The self-imaging phenomenon and its applications,
Prog. Opt. \textbf{27}, 1 (1989).

\bibitem{deng1999temporal}
L.~Deng, E.~W.~Hagley, J.~Denschlag, J.~E.~Simsarian, M.~Edwards, C.~W.~Clark, K.~Helmerson, S.~L.~Rolston, and W.~D.~Phillips,
Temporal, matter-wave-dispersion Talbot effect,
Phys. Rev. Lett. \textbf{83}, 5407 (1999).

\bibitem{ryu2006High}
C.~Ryu, M.~F.~Andersen, A.~Vaziri, M.~B.~d'Arcy, J.~M.~Grossman, K.~Helmerson, and W.~D.~Phillips,
High-Order Quantum Resonances Observed in a Periodically Kicked Bose-Einstein Condensate,
Phys. Rev. Lett. \textbf{96}, 160403 (2006).

\bibitem{rozenman2022periodic}
G. G. Rozenman, W. P. Schleich, L. Shemer, and A. Arie,
Periodic Wave Trains in Nonlinear Media: Talbot Revivals, Akhmediev Breathers, and Asymmetry Breaking,
Phys. Rev. Lett. \textbf{128}, 214101 (2022).

\bibitem{bakman2019observation}
A.~Bakman, S.~Fishman, M.~Fink, E.~Fort, and S.~Wildeman,
Observation of the Talbot effect with water waves,
Am. J. Phys. \textbf{87}, 38 (2019).

\bibitem{han2013wide}
C.~Han, S.~Pang, D.~V.~Bower, P.~Yiu, and C.~Yang,
Wide field-of-view on-chip Talbot fluorescence microscopy for longitudinal cell culture monitoring from within the incubator,
Anal. Chem. \textbf{85}, 2356 (2013).

\bibitem{Pfeiffer2008Xray}
F.~Pfeiffer, M.~Bech, O.~Bunk, T.~Donath, B.~Henrich, P.~Kraft, and C.~David,
X-ray dark-field and phase-contrast imaging using a grating interferometer,
IEEE Trans. Biomed. Circuits Syst. \textbf{105}, (2008).

\bibitem{bigourd2008factorization}
D.~Bigourd, B.~Chatel, W.~P.~Schleich, and B.~Girard,
Factorization of Numbers with the Temporal Talbot Effect: Optical Implementation by a Sequence of Shaped Ultrashort Pulses,
Phys. Rev. Lett. \textbf{100}, 030202 (2008).

\bibitem{pelka2018prime}
K.~Pelka, J.~Graf, T.~Mehringer, and J.~von Zanthier,
Prime number decomposition using the Talbot effect,
Opt. Express \textbf{26}, 15009 (2018).

\bibitem{liu2023axial}
X.~Liu, C.~Liang, Y.~Cai, and S.~A.~Ponomarenko,
Axial correlation revivals and number factorization with structured random waves,
Phys. Rev. Appl. \textbf{20}, L021004 (2023).

\bibitem{Iwanow2005Discrete}
R. Iwanow, D. A. May-Arrioja, D. N. Christodoulides, G. I. Stegeman, Y. Min, and W. Sohler,
Discrete Talbot Effect in Waveguide Arrays,
Phys. Rev. Lett. \textbf{95}, 053902 (2005).

\bibitem{zhang2014Nonlinear}
Y. Zhang, M. R. Belić, H. Zheng, H. Chen, C. Li, J. Song, and Y. Zhang,
Nonlinear Talbot effect of rogue waves,
Phys. Rev. E \textbf{89}, 032902 (2014).

\bibitem{schiek2021excitation}
R. Schiek,
Excitation of nonlinear beams: from the linear Talbot effect through modulation instability to Akhmediev breathers,
Opt. Express \textbf{29}, 15830 (2021).

\bibitem{ponomarenko2015self}
S.~A.~Ponomarenko,
Self-imaging of partially coherent light in graded-index media,
Opt. Lett. \textbf{40}, 566 (2015).

\bibitem{schouten2025gratings}
H.~F.~Schouten and T.~D.~Visser,
Gratings weave coherence carpets with non-diffracting coherence strands,
Optica \textbf{12}, 708 (2025).

\bibitem{Bai2026Regular}
Y. Bai, Y. Cai, C. Liang, and S. A. Ponomarenko,
Regular and irregular revivals of quasi-periodic random waves,
arXiv:2603.16371 (2026).

%\bibitem{liang2014ol}
%C. Liang, F. Wang, X. Liu, Y. Cai, and O. Korotkova,
%``Experimental generation of cosine-Gaussian-correlated Schell-model beams with rectangular symmetry,''
%Opt. Lett. \textbf{39}, 769--772 (2014).

%\bibitem{ma2019as}
%P. Ma, B. Kacerovská, R. Khosravi, C. Liang, J. Zeng, X. Peng, C. Mi, Y. E. Monfared, Y. Zhang, F. Wang, and Y. Cai,
%``Numerical approach for studying the evolution of the degrees of coherence of partially coherent beams propagation through an ABCD optical system,''
%Appl. Sci. \textbf{9}, 2084 (2019).

%\bibitem{liu2023oes}
%X. Liu, Q. Chen, J. Zeng, Y. Cai, and C. Liang,
%``Measurement of optical coherence structures of random optical fields using generalized Arago spot experiment,''
%Opto-Electron. Sci. \textbf{2}, 220024 (2023).

\bibitem{bai2025Self}
Y. Bai, Q. Chen, T. Cao, J. Liu, X. Liu, C. Liang, Y. Cai, and X. Peng,
Self-reconstruction properties of partially coherent pulses in nonlinear Kerr media,
Phys. Rev. A \textbf{112}, 033513 (2025).

\bibitem{chen2025Controllable}
Q. Chen, Y. Bai, X. Wang, P. Peng, J. Liu, Y. Cai, and C. Liang,
Controllable temporal dynamics of partially coherent pulses in nonlinear dispersive media,
Phys. Rev. A \textbf{112}, 033536 (2025).

\bibitem{ma2014optical}
L.~Ma and S.~A.~Ponomarenko,
Optical coherence gratings and lattices,
Opt. Lett. \textbf{39}, 6656 (2014).

\bibitem{OCL15T}
L.~Ma and S.~A.~Ponomarenko,
Free-space propagation of optical coherence lattices and periodicity
reciprocity, Opt. Express, \textbf{23} 1848 (2015).

\bibitem{OCL17T}
C. Liang, C. Mi, F. Wang, C. Zhao, Y. Cai, and S. A. Ponomarenko,
Vector optical coherence lattices generating controllable far-field beam profiles, Opt. Express, \textbf{25} 9872 (2017).

\bibitem{OCL16E}
Y. Chen, Y. Cai and S.A. Ponomarenko, Experimental generation of optical coherence lattices, Appl. Phys. Lett., \textbf{109} 061107 (2016).

\bibitem{OCL21E}
C Liang, X Liu, Z Xu, F Wang, W Wen, SA Ponomarenko, Y Cai, and P Ma, Perfect optical coherence lattices, Appl. Phys. Lett., \textbf{119} 131109 (2021).

\bibitem{PSA05}
S. A. Ponomarenko, H. Roychowdhury, and Emil Wolf, Physical significance of complete spatial coherence of optical fields, Phys. Lett. A, \textbf{345}, 10 (2005).

\bibitem{Akhmediev1987exact}
N.~N.~Akhmediev, V.~M.~Eleonskii, and N.~E.~Kulagin,
Exact first-order solutions of the nonlinear Schrödinger equation,
Theor. Math. Phys. \textbf{72}, 809 (1987).

\bibitem{dudley2009modulation}
J. M. Dudley, G. Genty, F. Dias, B. Kibler, and N. Akhmediev,
Modulation instability, Akhmediev Breathers and continuous wave supercontinuum generation,
Opt. Express \textbf{17}, 21497 (2009).

\bibitem{SI}
See Supplemental Material for the detailed derivation of the Fourier series expansion of the AB field and a supplementary video illustrating the dynamic evolution of the Talbot carpet for variable $\xi_c$.

\bibitem{Agrawal2012Nonlinear}
G. P. Agrawal,
\textit{Nonlinear Fiber Optics},
Academic Press (2012).

\end{thebibliography}

\end{document}